# Multiscale Investigation of Chemical Interference in Proteins

Antonios Samiotakis*[+], Dirar Homouz*[+] and Margaret S. Cheung*

* Department of Physics, University of Houston.

[+] These authors contributed equally to this work.

Corresponding author email: mscheung@uh.edu

**Abstract**

We developed a multiscale approach (MultiSCAAL) that integrates the potential of mean force (PMF) obtained from all-atomistic molecular dynamics simulations with a knowledge-based energy function for coarse-grained molecular simulations in better exploring the energy landscape of a small protein under chemical interference such as chemical denaturation. An excessive amount of water molecules in all-atomistic molecular dynamics simulations often negatively impacts the sampling efficiency of some advanced sampling techniques such as the replica exchange method and it makes the investigation of chemical interferences on protein dynamics difficult. Thus, there is a need to develop an effective strategy that focuses on sampling structural changes in protein conformations rather than solvent molecule fluctuations. In this work, we address this issue by devising a multiscale simulation scheme (MultiSCAAL) that bridges the gap between all-atomistic molecular dynamics simulation and coarse-grained molecular simulation. The two key features of this scheme are the Boltzmann inversion and a protein atomistic reconstruction method we previously developed (SCAAL). Using MultiSCAAL, we were able to enhance the sampling efficiency of proteins solvated by explicit water molecules. Our method has been tested on the folding energy landscape of a small protein Trp-cage with explicit solvent under 8M urea using both the all-atomistic replica exchange molecular dynamics (AA-REMD) and MultiSCAAL. We compared computational analyses on ensemble conformations of Trp-cage with its available experimental NOE distances. The analysis demonstrated that conformations explored by MultiSCAAL better agree with the ones probed in the experiments because it can effectively capture the changes in side chain orientations that can flip out of the hydrophobic pocket in the presence of urea and water molecules. In this regard, MultiSCAAL is a promising and effective sampling scheme for investigating chemical interference which presents a great challenge when modeling protein interactions *in vivo*.

# I. INTRODUCTION

Proteins are the molecular "workhorses" that carry out functions in living organisms. Proteins need to fold into well defined compact structures in order to perform these functions. Such "protein folding" events depend on the amino acid sequences of proteins and their surrounding environment. The physical principle of this process in a crowded and confined cell, however, has not been fully understood[1]. The major obstacles lie in complex interactions between odd-shaped macromolecules which often chemically interfere with one another in a crowded space, making experiments extremely difficult. This problem could be better understood by utilizing computer simulation and modeling that offer an effective approach to explore a broad range of parameters and solvent conditions, reducing costs to experiments[2-8]. However, performing such a task is challenging because it involves computer simulation of a biological system whose spatiotemporal dynamics span across multiple orders of magnitude that is beyond the capacity of current computing resources.

The recently advanced computer simulations remain in all-atom and coarse-grained molecular simulations which have different strengths as well as shortcomings. All-atomistic molecular dynamics simulations with explicit solvent molecules provide a great deal of detail of protein dynamics on a very short-time scale (e.g. picoseconds). However, it can be too computationally costly[9] to produce a meaningful folding trajectory that typically elapses in microseconds. On the other hand, Coarse-grained (CG) protein models are very effective in capturing the main features of a protein and have the ability to simulate full protein folding trajectories up to several microseconds[10]. However, CG protein models lack the atomistic details and reliable energy functions that reflect changes in solvent conditions, so they appear "not realistic" enough[11]. Because of these shortcomings, CG investigations have been restricted to isolated proteins in aqueous solutions. Other methods, like the Generalized Born method[12, 13], attempt to reduce the computational cost by treating the solvent implicitly. However, these approaches are also restricted in aqueous solutions and have yet to address the issue of chemical interference.

To solve more complex problems with larger proteins over a broad range of spatiotemporal scales, in terms of multiple orders of magnitudes, several approaches that can mix CG and all-atomistic (or fine-grained, FG) protein models are being developed. Some multiscale algorithms bridge the length scales of FG and CG models. A "force matching" technique that gathers forces on CG models from FG simulations[14, 15] has been developed and applied to simple systems[16]. A hybrid FG/CG molecular simulation system in which the resolutions can be dynamically switched has been proposed[17-20]. Alternatively, some methods take hydrodynamic interactions into account and integrate disparate time scales between molecular dynamics and Brownian dynamics[21]. However, these approaches are only applicable to simple isolated systems without complex changes in solvent environment; fundamental concerns of how to integrate force fields from different scales still remain elusive. Little is known, for example, of how to take radical chemical interferences into account on the self-assembly of macromolecules. Therefore, an investigation of protein dynamics and the folding energy landscape inside a cell, a crowded medium where interactions between proteins are often changed by chemical interference, is not available; thus, development of multi-scale approaches to serve this purpose of prompt integration of force fields for protein models at different resolutions is urgently needed. In addition, developments of new efficient multiscale modeling algorithms are required to optimize data transmission that enhances the computing capacity for all system sizes.

In this work, we developed a multi-scaled molecular simulation method (MultiSCAAL) and used it to construct a well-sampled folding energy landscape of a Trp-cage protein under an aqueous condition and at high level of urea concentration that would otherwise be very difficult to simulate using the approach of molecular dynamics simulation alone. Trp-cage is a small fast-folding protein with 20 amino acids that has been studied extensively by experiments[22-24] and it is often used to gauge the validity of force fields[25] and new computational methods[26-31]. Our method is built on integration of several well-established approaches such as coarse-grained protein models[32, 33] (Side Chain $C_\alpha$ Model, SCM), reconstruction of all-atom protein models from SCM models[6] (SCAAL), and all-atomistic molecular dynamics simulations[34, 35]. Our strategy is composed of three steps: (1) The energy function for coarse-grained molecular simulation is derived from the potential of mean force (PMF) from the all-atomistic simulations that contain certain chemical interference using Boltzmann inversion method[36, 37]; (2) Coarse-grained protein representations in a thermodynamic ensemble of interest are selected according to a Metropolis criterion[38] and all-atomistic protein conformation are promptly reconstructed by effectively incorporating an all-atomistic protein model as a template; (3) Folding free energy landscape of a protein that uses reconstructed all-atomistic protein models built from step (2) as initial conformations is effectively simulated by all-atomistic molecular dynamics. A schematic overview of the algorithm is presented in Fig.1 and the focus of our study in part lies in the transition between the energy function for coarse-grained models and the PMF.

While the method of Boltzmann inversion has been applied in other studies[36 39], the main issue that limits its practical use is the justification of a reliable reference state to relate the non-bonded energy function of a CG protein model to the PMF obtained from all-atomistic molecular simulations[40]. Here, we attempted to resolve this problem by using an energy function taken from a matrix of knowledge-based potential (or statistical potential) determined from the Protein Data Bank[41] and tested our approach on a small protein, Trp-cage. The rationale is the following: given that statistical potentials for coarse-grained molecular simulations and the force fields for all-atomistic molecular simulations have been independently derived from different experiments, both have been successfully applied for the investigation of research topics that address protein dynamics and structural interactions. By establishing an empirical relationship between PMF from all-atomistic systems and the energy function for non bonded interactions in coarse-grained systems, we may be able to apply this knowledge for an effective integration of multiscale molecular simulations as well as to improve the exploration of the folding energy landscape.

We constructed a folding energy landscape of a small protein Trp-cage under an aqueous condition and 8M urea concentration using all-atomistic replica exchange molecular dynamics (AA-REMD) and MultiSCAAL. Selected atomistic distances in dominant ensemble structures of Trp-cage using both methods were compared with NOE results from experimental NMR studies [24]. Under the aqueous condition both methods produce nearly the same dominant protein structures. Interestingly, under 8M urea conditions MultiSCAAL was able to render a broader energy landscape of Trp-cage than AA-REMD by providing a wider variety of protein conformations in which side chains are able to flank out of the hydrophobic pocket in the presence of urea. The result of a flipped tryptophan captured by MultiSCAAL better explained short atomistic distances to neighboring amino acids indicated by NOE measurements[24].

## II. THEORY AND METHODS

## 1. ENERGY FUNCTION DEPENDENT ON CHEMICAL INTERFERENCE

### 1.1 Boltzmann inversion

An energy function was developed for coarse-grained molecular simulation that accommodates chemical interference using Boltzmann inversion[37, 42, 43] from the data derived from all-atomistic molecular dynamics simulations. The pair correlation function between any two amino acid types $i$ and $j$ at a distance $r$ in solvent type $\alpha$ is $g_{ij}^{\alpha}(r)$. $r^*$ denotes the first highest peak of $g_{ij}^{\alpha}(r)$, where $g_{ij}^{\alpha}(r^*)$ is a maximum. $g_{ij}^{\alpha}(r)$ relates to the potential of mean force, $U_{ij}^{a}(r)$, between the same pair of amino acids through Boltzmann inversion at temperature T by the following formula[36] :

$$U_{ij}^{\alpha}(r) = -k_B T \ln\left[\frac{g_{ij}^{\alpha}(r)}{\rho_o}\right] \qquad \text{eqn 1}$$

where $\rho_o$ is the average density of the system (amino acid pairs and the solvent) and $k_B$ is the Boltzmann constant. The average density $\rho_o$ is used to normalize the pair correlation function at distances, greater than the excluded volume radius. The solvent mediated interactions $\varepsilon_{ij}^{'\alpha}$ for every pair of amino acids $i$ and $j$ is $U_{ij}^{\alpha}(r^*)$. We next shift $\varepsilon_{ij}^{'\alpha}$ by a constant, $V_o$ ,

$$\varepsilon_{ij}^{\alpha} = \varepsilon_{ij}^{'\alpha} + V_o \qquad \text{eqn 2}$$

where $V_o$ is obtained from a Threonine pair by setting $\varepsilon_{TT}^{'\alpha}$ from the simulation equal to $\varepsilon_{TT}^{\alpha}$ from the statistical potential of the same amino acid pair[41].

A Lennard-Jones potential (LJ), $V_{ij}^{a}(r)$, was used to approximate the overall profile of $U_{ij}^{a}(r)$ [44] and it is the energy function for the same type of amino acids in coarse-grained molecular simulation:

$$V_{ij}^{\alpha}(r) = \varepsilon_{ij}^{a}\left[\left(\frac{r_{ij}^{o}}{r}\right)^{12} - 2\left(\frac{r_{ij}^{o}}{r}\right)^{6}\right] \qquad \text{eqn 3}$$

$\varepsilon_{ij}^{\alpha}$ is the solvent-mediated interaction of an amino acid pair $i$ and $j$ in solvent type $\alpha$. $r_{ij}^{o}$ is the bonding distance which will be further elaborated in Section III.1

## 1.2 Computation of pair correlation functions from all – atomistic molecular dynamics simulations with explicit solvent molecules

In order to fit the parameters of solvent-mediated interaction, $\varepsilon_{ij}^{\alpha}$, we computed the correlation functions of pairs of amino acids in ~2M concentration for each type of amino acid solvated in explicit water. Amino acids were represented by the United Atom Model of the amber force field ff03ua[45] without caps (as it was the same construct in another study[36]), and the water molecules were represented by the TIP3P model[46]. Sodium and chlorine ions (Na$^{+}$, Cl$^{-}$) were added accordingly, in order to neutralize systems with charged groups. The cutoff for non-bonded interactions is 1nm. Electrostatic interactions were treated with the Particle Mesh Ewald method (PME)[47]. The direct space cutoff for electrostatic interactions is 1nm, while the Fourier space part was computed using a grid spacing of 1Å and cubic spline interpolation. A periodic cubic box of a fixed volume with a side L = 40Å was used for initial preparation of the system.

The system went through several preparation steps before the production run. First, energy minimization was performed using the conjugate gradient method for 1000 steps. Then, each system was heated to 300K in an incremental temperature step of 100K under NVT condition for 50ps. To achieve adequate water density, we performed another equilibration step at the constant pressure of 1.0 atm and of temperature 300K (NPT), maintained by the Andersen algorithm[48]. Finally, another 50 ps constant volume and temperature (NVT) equilibration was performed prior to the following NVT production run. In the production run, a 50 ns NVT simulation was performed.

To extract $g^{\alpha}{}_{ij}(r)$ of each amino acid pair $i$ and $j$ from the simulation, we created a histogram from the distances among the different amino acid pairs with a bin size of 0.1Å. The distance between two amino acids is measured from the heavy atom that is closest to the center of mass of a side chain (Table 1) of each amino acid. This procedure was performed for all 210 combinatorial amino acid pairs, yielding a matrix of interactions $\varepsilon'^{aq}_{ij}$.

We repeated the same procedures to extract the solvent mediated interactions, $\varepsilon''^{ur}_{ij}$, for each amino acid pair in the presence of 8M urea. The urea molecules are represented by the same AMBER force field as the one used for the amino acids.

## 2. COARSE-GRAINED MOLECULAR SIMULATION ON TRP-CAGE

Coarse-grained (CG) molecular simulations were performed on a mini protein Trp-cage to investigate its folding energy landscape. Trp-cage has 20 amino acids (PDBID: 1L2Y) and it is represented by a side-chain $C_{\alpha}$ model (SCM), in which each amino acid is represented by two beads (excluding glycine) (see the Method Section in the Supplement). The stochastic dynamics to represent the solvent effects is represented by Langevin equations of motion[49] (see the Method

Section in the Supplement). In the aqueous solution (in absence of chemical interference) the nonbonded interactions follow the Betancourt-Thirumalai statistical potential[41] in which threonine-threonine interaction was used as the reference to fit hydrophobic parameters obtained from partition chromatography experiments[50]. In the 8M urea solution, the nonbonded interactions were taken from the Boltzmann inversion as described in Section 1.1.

In the production run, the Replica Exchange Method (REM)[51] was applied to enhance sampling efficiency by incorporating multiple copies (replicas) of molecular simulation at a broad range of temperatures. Exchanges between neighboring replicas $i$ and $j$ are accepted with a probability:

$$P_{acc} = \min\left\{1, \exp\left[\left(\beta_i - \beta_j\right)\cdot\left(U(r_i) - U(r_j)\right)\right]\right\} \quad \text{eqn 4.}$$

where $\beta = 1/k_BT$ and $U(r)$ represents the potential energy of the system. 16 replicas are distributed between 280K and 580K. An in-house version of AMBER6[52] is used, where the Langevin equations of motion are integrated in a low friction limit. Each replica produces ~200,000 statistically significant conformations sampled with time separations greater than one characteristic correlation time [53].

# 3. RECONSTRUCTION OF ALL-ATOMISTIC PROTEIN STRUCTURES FROM COARSE-GRAINED MODELS

## 3.1 Reconstruction algorithm: Side-Chain C$_\alpha$ model to All-atom (SCAAL)

In order to reconstruct a desired all-atomistic structure from its coarse-grained SCM model, its bead positions were used as a part of harmonic constraints and applied to an all-atomistic protein template in the energy minimization step, using the conjugate gradient method.

The prototype of SCAAL for the reconstruction of several compact structures was introduced in our prior study[6]. For each residue, $C_\alpha$ positions from the SCM were used as position constraints for $C_\alpha$ in the backbones from the all-atomistic template. The constraints on a side chain were imposed on a heavy atom with the closest proximity to the center of mass of the side chain which is typically less than 1Å. Selection of the chosen side chain heavy atom for each amino acid is listed in Table 1.

Next, through the energy minimization, harmonic constraints imposed by the chosen beads bring the all-atom template to the desired structure without the need for building a protein from individual atoms. The spring constants $k_\alpha = 1000 kcal/mol \cdot \text{Å}^2$ are used for the constraint applied on a $C_\alpha$ bead, while for a side chain bead is $k_b = 40 kcal/mol \cdot \text{Å}^2$. These values are derived after extensive testing to ensure that the reconstructed structures are accurate and do not experience steric frustrations. The concept of using an atomistic "template" for protein reconstruction is depicted schematically in Fig. 2a and the flowchart of the SCAAL reconstruction algorithm is illustrated in Fig. 2b.

*Reconstruction of unfolded conformations from coarse-grained protein models:* To test our reconstruction algorithm, a pool of 67 non–redundant tested proteins was selected from the protein data bank using the same criteria as in Feig et al[54] that is exclusively determined by X–ray diffraction experiments (resolution of 1.00 Å or better). The pdb ID's of these structures are provided in the Supplement. Unfolded conformations of each tested protein are generated at high temperatures. To ensure such unfolded conformations are of little resemblance to the minimized native structures, we set the criteria of structural selection for unfolded structures where the RMSD (root-mean-square-deviation) of heavy atoms with respect to the minimized native structure exceeds 10Å and its radius of gyration is at least 15% greater than the minimized native state. Each unfolded conformation is coarse grained into an SCM model (keeping $C_\alpha$ and the center of mass of side chains) and reconstructed into all-atomistic protein models using SCAAL.

## 4. ALL-ATOMISTIC MOLECULAR SIMULATIONS FOR THE FOLDING ENERGY LANDSCAPE OF TRP-CAGE

In this study, we constructed the folding energy landscape of Trp-cage using two different methods: (1) Using the method of MultiSCAAL we developed in this work (Section 4.1), and (2) using the standard all-atomistic molecular dynamics with the Replica Exchange Method (AA-REMD) as described in Section 4.2.

### 4.1 MultiSCAAL

The purpose of MultiSCAAL is to enhance the sampling of all-atomistic molecular dynamics simulations by selecting a large set of initial conditions selected from the CG distributions. This scheme allows all atomistic molecular dynamics simulation, following all-atomistic force fields, to efficiently probe and refine the conformations that are strategically reconstructed from CG protein models. Our scheme is distinct from the concept of Resolution Exchange [55, 56] that performs iterative conformation exchanges between the CG and AA molecular simulations. Instead, we concentrate on the proper selection of initial AA conformations based on CG protein model with knowledge-based energy function that can be adjusted to different environmental conditions as detailed below. First, the conformations sampled from the ensembles of CG molecular simulations in the presence and absence of urea were selected based on the Metropolis criterion[38]. The potential energy of the most populated configuration (global minimum in the free energy) in the CG ensemble is $E_{ref}$, and the energy of state $i$ that is randomly chosen from the CG ensemble is $E_i$. The acceptance probability $P_{acc}$ at temperature T is defined as following:

$$P_{acc} = \min(1, \exp(-\Delta)), \text{ with } \Delta = \frac{E_i - E_{ref}}{k_B T}, \quad \text{eqn. 5}$$

$k_B$ is the Boltzmann constant. One in every 1000 conformations was randomly sampled from the simulation data and subsequently reconstructed into all-atomistic representations by SCAAL as initial conditions for subsequent all–atomistic molecular dynamics simulations.

Secondly, a reconstructed all-atomistic protein model was solvated by explicit water molecules in a periodic cubic box. Depending on the size of a Trp-cage, the sizes of each water/urea box were different for each conformation, ranging between 40Å to 45Å. If the structure was reconstructed from the coarse-grained molecular simulation with the energy function for urea condition, then an 8M urea box is used for solvation. Finally, after solvating the system, a short equilibration was run in a constant volume and temperature of 300K. All-atomistic molecular dynamics simulations of Trp-cage were run at 300K for 1-ns.

## 4.2 All-atomistic molecular dynamics with replica exchange method (AA-REMD)

A set of AA-REMD was performed on Trp-cage for aqueous conditions and with 8M urea, respectively. Simulations were performed at constant volume and constant temperature (NVT) conditions, in a periodic cubic box as in similar studies[27]. The AMBER force field ff99SB[57] was used for Trp-cage, and urea and water molecules were represented by the TIP3P model[46]. Each system initially underwent energy minimization using a combination of steepest descent and conjugate gradient steps, followed by heating to 300K under NVT conditions with a temperature step of 100K and then equilibrated at 300K in NPT conditions to adjust the correct solvent density in each system. Finally a set of short equilibration runs of 0.1ns is performed under NVT conditions in order to create the different initial conditions for the Replica Exchange simulations. For each system, different replicas are equilibrated at 40 temperatures between 280 K and 540K. Replica Exchange Method (REM) simulations were applied using these 40 different replicas, each with different initial conditions, generating ~50,000 different configurations from 5,000 exchanges where each exchange is attempted at 2,000 steps. The fraction of exchanged replicas over the attempted ones (exchange rate of REM) is ~20%.

## III. RESULTS AND DISCUSSION

## 1. Relationship of the statistical potential and the energy function derived from the potential of mean force

One of the important steps for multiscale molecular simulation is to bridge the energy functions between protein models with different resolutions. There is a need to develop a reliable algorithm that can accurately transfer essential information such as chemical interference from high-resolution model to low-resolution model without losing its key features. In this regard, we adopted the method of Boltzmann inversion (eqn.1) that provides a link between the potential of mean forces derived from the all-atomistic molecular dynamics simulation and the energy function for the coarse-grained molecular simulation, which has been applied in other studies [36, 37, 39]. However, computation of the potential of mean force from all-atomistic molecular dynamics simulation is inherently complex and inefficient. In addition, it is questionable whether the thermodynamic properties of coarse-grained molecular simulations can be accurately transferred by the solvent-mediated interaction derived from the potential of mean force. Here, we developed a new and simple approach using the statistical potential parameter table (SPPT) to resolve these issues.

We provided a method to calibrate a reference state of the energy function for coarse-grained molecular simulation (eqn.2) by incorporating statistical potentials from prior bioinformatics studies (eqn.3). Although the interaction parameters for all-atomistic molecular dynamics simulations and the statistical potentials for coarse-grained molecular simulations are both derived from different sets of experimental data, the two should be correlated and this relationship can be used to bridge the molecular simulations in different resolutions. Using this relationship, we can further develop a statistical potential parameter table (SPPT) of amino acid interactions with chemical interference from all-atomistic molecular dynamics simulations with corresponding chemical conditions.

To determine the relationship of the two energy functions, first we performed all-atomistic molecular simulations of each amino acid pair in aqueous conditions and computed its pair correlation function $g_{ij}^{aq}(r)$. Using a pair of threonine-threonine (TT) as an example, we calculated its pair correlation function, $g_{TT}^{aq}(r)$ shown in Fig. 3a. After Boltzmann inversion, the solvent-mediated interaction, $\varepsilon_{TT}^{'aq}$, was obtained from its potential of mean force $U_{TT}^{aq}(r^*)$ (Fig. 3b), where $r^*$ locates the major peak of $g_{TT}^{aq}(r)$ in Fig. 3a. All of the possible 210 solvent-mediated interaction $\varepsilon_{ij}^{'}$ between a pair of amino acid $i$ and $j$ in aqueous condition were obtained from the same framework as the one described above.

Next, we plotted $\varepsilon_{ij}^{'aq}$ against the energy function from the Betancourt-Thirumalai statistical potential[41] of the same amino acid pair, $\varepsilon_{ij}^{aq}$ in Fig. 3. The two sets of parameters correlate very well with a linear regression coefficient 0.79. The distributions of $\varepsilon_{ij}^{'aq}$ and $\varepsilon_{ij}^{aq}$ are also similar (See Supplement Fig S1). The vertical shift of $\varepsilon_{ij}^{'aq}$ can be offset by matching Thr-Thr interaction in both parameter sets with $V_o = 0.31 kcal/mol$. This conceptual paradigm was expanded from the Betancourt-Thirumalai study in which the statistical potentials of Miyazawa-Jernigan[58] were shifted using Thr-Thr interaction as a reference state to match experimental studies. By doing so, we established a reliable tool in transferring the energy functions in all-atomistic and coarse-grained molecular simulations, in which both parameters were tested against different experimental factors.

In a Lennard-Jones potential $V_{ij}^{aq}(r)$ (eqn.3) for coarse-grained molecular simulation, $r^o{}_{ij}$ is the bonding distance between a pair of amino acid $i$ and $j$. The potential was obtained by the measurement of the distance between the centers of mass of two interacting amino acids. The observed discrepancy between $r^*$ and $r^o{}_{ij}$ (Fig. 3b) is due to the fact that $r^*$ was sampled from amino acid pairs with uneven atomistic structures, while $r^o{}_{ij}$ was computed assuming a smooth spherical side chain. This difference exists in all amino acid pairs, but it is not dependent on solvent conditions, therefore, we simply keep $r^o{}_{ij}$ in coarse-grained molecular simulation for the energy function in Fig. 3c.

Once we established a reliable relationship of the energy functions between all-atomistic and coarse-grained molecular simulation, we were able to repeat the same procedure on amino acid pairs in the presence of 8M urea as the aqueous condition in order to obtain SPPT under the chemical inference of 8M urea. The solvent-mediated interactions, $\varepsilon_{ij}^{'ur}$, derived from the potential of mean force, were offset to $\varepsilon_{ij}^{ur}$ for the coarse-grained energy function by the same $V_o$ as the one used for the aqueous condition. $\varepsilon_{ij}^{ur}$ for all pairs of amino acid $i$ and $j$ in the presence of 8M urea is listed in Table S1. $\varepsilon_{ij}^{ur}$ is overall less than $\varepsilon_{ij}^{aq}$ and reflects weakened interactions of amino acid pairs in the presence of 8M urea, leading to protein destabilization. The contact energies of the Betancourt-Thirumalai statistical potential are provided in Table S2 as reference.

## 2. Coarse-grained molecular simulation on Trp-cage under urea condition

With an energy function that can incorporate different conditions, we produced the folding energy landscape of Trp-cage using coarse-grained molecular simulations for both aqueous and 8M urea conditions (Fig. 5 a, b respectively). The free energy landscape at 300K is plotted as a function of the root-mean-square-deviation (RMSD), which provides a quantitative measure of conformational changes and the radius of gyration ($R_g$) that determines the size of a protein.

In Fig.5a, the folded Trp-cage in water is in the region of $2.0\text{Å} \leq RMSD \leq 3.5\text{Å}$ and $7.3\text{Å} \leq R_g \leq 8.0\text{Å}$. Under high concentration of urea in Fig.5b, the main free energy basin is in the region of $2.0\text{Å} \leq RMSD \leq 3.5\text{Å}$ and $7.7\text{Å} \leq R_g \leq 8.5\text{Å}$. While the main basin shifts to higher values of $R_g$ in the presence of 8M urea, the population of structures with low $R_g$ decreased. In addition, the population of structures with $R_g > 8.5\text{Å}$ is enhanced in the presence of denaturant conditions. We used a self-organizing neural net[59, 60] clustering technique[61, 62] to distinguish the structures without setting any prior structural assumptions.

Typical representative conformations sorted by the clustering method from the two basins are presented for water (Fig.6a) and for 8M urea (Fig.6b). The dominant structure found in the aqueous coarse–grained molecular simulations is similar to the NMR structure obtained from experiments[24]. The radius of gyration of this ensemble is ~7.6Å which is very similar to that of the NMR structure of Trp-cage (~7.5 Å). The position of Trp6, as shown in Fig.6e, illustrates that the hydrophobic core of Trp-cage remains stable throughout the simulations.

In contrast, under 8M urea conditions the coarse–grained molecular simulations demonstrated that the size of the radius of gyration of the most dominant cluster grows ($R_g \sim 8.2$ Å), compared to the aqueous simulations. In this state, the position of Trp6 is no longer "caged" by the hydrophobic residues of the protein (Tyr3, Leu7 Pro12, Pro17, Pro18, Pro19) as shown in Fig. 6b. This agrees with previous theoretical studies of Trp-cage that suggest a similar mechanism of unfolding for Trp–cage, where the structure does not unfold to a random-coil conformations but rather increases in size and exposure of the hydrophobic residues[25, 26].

### 3. Reconstruction of all-atomistic Trp-cage structure from coarse-grained protein models

The accurate reconstruction of high-resolution protein models from low-resolution ones is a necessary step to incorporate chemical interferences in a multi-scale molecular simulation. Several strategies based on the search for a globally minimized state are used for reconstructing reduced protein to all-atomistic structures over the last years[63-65]. Although a strong dependence on rotamer libraries in some of these algorithms has enabled fast and accurate reconstruction methods for the native state protein structures, it is questionable whether the accuracy of such methods can still be achieved or not for protein conformations far from their native states. This aspect is particularly important in the development of a multiscale molecular simulation approach in which protein structures produced at a condition overwhelmingly different from a globally minimized state (e.g. crystal structures). Without an accurate method for protein reconstruction in unfolded conformations, the validity of an algorithm for multiscale molecular simulation is jeopardized.

We tested the accuracy of reconstruction of unfolded proteins conformations using SCAAL (See Method section) against PULCHRA[63], in Fig.7. Both of the two methods require a $C_\alpha$ bead and the center of mass of side chain as input. However, the major difference lies in the fact that PULCHRA requires a side-chain rotamer library, which is obtained from the protein data bank for the reconstruction of side chains, while SCAAL uses a randomly selected all-atomistic protein conformation of a tested protein as a template for side chain reconstruction.

We produced unfolded protein structures (See Method section) using high temperature molecular simulation models, coarse-grained them into SCM low-resolution protein models and reconstructed them back to all-atomistic protein models using SCAAL and PULCHRA. We then compared the RMSD between these reconstructed structures and the original chosen unfolded atomistic structures on a basis of residue types as an indicator for the accuracy of the two reconstruction methods in Fig.7. An overall performance regarding accuracy undertaken by SCAAL is much better than PULCHRA by 30%. The average RMSD constructed by SCAAL is within 1.33 Å, and it worked better on amino acids with larger side chains (e.g. R, H, W, K, F, Y, and E) than PULCHRA. This suggests that a reconstruction of side chains based on physical models using harmonic constraints will be more reasonable than the incorporation of rotamer libraries (based on information of native folded states) as far as the reconstruction of unfolded states is concerned.

## 4. Free energy landscape of Trp-cage produced from all-atomistics molecular simulations with REM and MultiSCAAL

A 2D-free energy landscape as a function of RMSD and $R_g$ of Trp-cage produced by all-atomistic molecular simulations with the replica exchange method (AA-REMD, see Method section) at 300K under aqueous and 8M urea is shown in Fig. 8 (a, b), respectively. In Fig.8a, there is a major population of conformation in aqueous solvent characterized by RMSD~2.5Å and Rg~7.5Å. In Fig.8b, the position of the major population shifts to RMSD~3.5 Å and Rg~8.0 Å in the presence of 8M urea, indicating that the population of swelled conformations induced by urea increases compared to the ones in water.

Using the same RMSD and $R_g$ parameters, a 2-D free energy landscape is also produced by all-atomistic molecular simulation with MultiSCAAL at 300K, under aqueous and 8M urea in Fig.8 (c, d) (See Method Section). In comparison to Fig.8(a, b), the area of free energy landscape explored by MultiSCAAL is broader in Fig.8(c, d), indicating enhanced sampling for both Trp-cage systems (in water and urea). Even though the position of the main basins in the free energy landscape are similar, the structural characteristics and ensemble distribution of Trp–cage sampled by the MultiSCAAL scheme contains a wider variety of structures than the ones in the AA-REMD simulations as indicated by greater value of RMSD. To further investigate the differences in the structural distribution in the free energy landscape obtained by the two methods (e.g. AA-REMD and MultiSCAAL) we used the clustering technique to distinguish the structures without setting any prior structural assumptions. The most dominant structures obtained from each method, under different conditions, are shown in Fig. 6 for visual comparison.

In the following sections, we investigated dominant ensemble conformations of Trp-cage in aqueous environment and 8M urea in detail:

(a) *Aqueous environment:* We used the clustering method [61, 62] to characterize the three most dominant structures in aqueous conditions sampled by AA-REMD and by MultiSCAAL as shown in Fig.9 at 300K. The first most dominant structures in the folded state with Rg~7.4 Å (Fig.9 a & d) obtained using both approaches are similar given that the calculated RMSD of the heavy atoms is <1.0Å. Using a characteristic hydrogen bond between the hydrogen atom (HE1) of W6, and the oxygen atom (O) of R16 (bond W6-R16) that is unique in the folded state of Trp-cage (determined by NMR[22]), we evaluated the dominant clusters of the folded states obtained from AA-REMD and MultiSCAAL by the presence of this hydrogen bond in clusters. For the AA-REMD, this hydrogen bond is present in 75% of the structures in the most dominant cluster and for the MultiSCAAL it is 78.3%. This result further supports the resemblance of the two most dominant clusters obtained from the two enhanced sampling techniques in an aqueous environment and they agree with experimental findings.

In the second and third most dominant clusters obtained with both methods in aqueous solvent (Fig.9 b & c), we find that the structures from AA-REMD simulations are similar to the ones belonging to the most dominant cluster, ranging from 7.36 Å–7.49 Å. However, in the case of the MultiSCAAL simulations, while Trp–cage retains its folded conformation, the structures differentiate in the position of the side chains of the main hydrophobic residues in Fig.9(e & f) resulting in a richer population of structures with similar $R_g$ but higher values in RMSD due to a wide range of orientation of side chains. Sampling all these changes in different orientations is a very difficult problem in all-atomistic molecular simulations with explicit water models.

(b) *8M urea:* In the dominant clusters, the unfolded states of Trp–cage obtained from all atomistic molecular simulation with REM (AA-REMD) and MultiSCAAL differ in the extent of packing of Trp6 against prolines (Fig.10 a & d). For both dominant clusters the radius of gyration increases to $R_g \approx 8.10$ Å in the presence of 8M urea which agrees with previous studies [24, 25] where an increase of ~10% of the radius of gyration induced by 6M urea was observed. Using the aforementioned W6-R16 characteristic hydrogen bond that is unique to the folded state of Trp-cage, we found that its presence is only in 0.03% and 0.08% of the dominant structures obtained by AA-REMD and MultiSCAAL, respectively (Fig.10 a & d).

Although the sizes of the conformations generated by MultiSCAAL and AA-REMD are similar, the structural details are quite different. From the conformations obtained by the MultiSCAAL approach, the sidechain of Trp6 (an indole group) can completely flip outward and exit from the hydrophobic core of a Trp-cage protein as illustrated in Fig.10d. In addition, when comparing the second and third most dominant clusters from both methods, we can see that in the case of REM, the sampled structures are similar, with $R_g$~8.09 Å and $R_g$~8.10 Å respectively (Fig. 10 b & c). In contrast, the structures in the second and third dominant clusters from the MultiSCAAL simulations are different with $R_g \sim 7.89$ Å and $R_g \sim 8.20$ Å, respectively (Fig.10 e & f). In order to evaluate the quality of these structures, we turn to a direct comparison with available experimental data in the next section.

## 5. Comparison with NMR NOE distances

In a recent experimental NMR study of Trp-cage at 6M urea[24], NOE distances were determined between the protons of the side chain (indole group) of Trp6 and the following amino acids: Ile4, Leu7, Pro12 and Arg16. It was found that under high concentration of urea and at low temperature (278K) the majority of these NOE distances were shorter than the ones in the native state of Trp–cage, even though the radius of the molecule as obtained from diffusion experiments[24] was found to be ~8Å, that is not far from the size of Trp-cage in aqueous condition. In this regard we calculated the NOE distances between the same proton pairs as in these experiments, for the three most dominant clusters obtained from both AA-REMD and MultiSCAAL simulations under 8M urea condition (Table 2) as a gauge to validate the accuracy of the folding energy landscape of Trp-cage under high content of urea obtained by two different methods. Although the urea concentration in our study is slightly greater than the one used in the experiments, it is still representative of a Trp-cage under high concentration of urea.

In the case of the clustered Trp-cage structures obtained from AA-REMD the majority of the NOE distances greatly deviate from the experimental values, as seen in Table 2. In contrast, the computed NOE distances of the dominant cluster of the Trp-cage structures obtained from MultiSCAAL exhibit remarkable similarity to the one measured by the NMR experiments, in spite of the small difference in the concentration of urea between the experiment and simulations. Ensemble structures sampled by MultiSCAAL match the distance constraints of the NMR experiments better than the ones obtained by AA-REMD, indicating the improved sampling in MultiSCAAL. Examining the Trp-cage structure in the dominant cluster from the MultiSCAAL simulation, we determined that a key factor that contributes to the shortening of the NOE distances is the exit of the side chain of Trp6 outside of the hydrophobic core (Fig. 11). This topological arrangement reduces the distances between proton in Trp6 (HE3) and Ile4 (HG2) to 6.1 Å, which is much closer to the experimental value (4.5 Å), than the distance of 8.0 Å in the ensemble structures sampled by AA-REMD simulations.

We further investigated the reason behind this topological arrangement of the Trp6 by measuring the number of water and urea molecules that are present in the hydrophobic pocket of the protein (e.g. by following the water and urea molecules shared by amino acids Trp6 and Arg16) In the average ensemble structures of Trp-cage in the dominant cluster sampled by AA-REMD, there are 0.53 water molecules and 1.94 urea molecules that appear in the hydrophobic pocket while in the average ensemble structure of Trp-cage in the dominant cluster sampled by MultiSCAAL there are 0.97 water molecules and 1.86 urea molecules in the hydrophobic pocket. This suggests, that there is a better chance for the presence of water molecules inside the hydrophobic core when the side chain of Trp6 (indole group) points away from the core as being adequately captured in the MultiSCAAL simulation and this structural feature can better account for the short NOE distance between Trp6 and other amino acids. Recently, it was demonstrated in a computational study[25] that in order to match the three NOE distances between Ile4 and Trp6 from NMR experiments[24] under 6M urea at 278K, the system had to be heated to 360K where the unfolding procedure could be expedited. Because MultiSCAAL incorporates coarse-grained protein models with a fewer number of side chain beads, it is easier to use coarse-grained models for exploring different orientations of side chain positions. This avoids the high entropic cost of rearranging solvent water molecules around side chains when Trp6 exits the hydrophobic core, so that sampling efficiency of protein conformations solvated by explicit water molecules can be greatly enhanced. We believe that it is for this reason that the structures sampled by MultiSCAAL simulation scheme can better match with the experimental NOE distances at a temperature closer to the experimental condition.

## 6. Overall algorithm performance of MultiSCAAL

We benchmarked the performance of the MultiSCAAL against AA-REMD by measuring the computational time in terms of CPU hours (CH) on 160 CPUs for each AA-REMD simulation on a Linux cluster (AMD Opteron 2.3GHz) at the University of Houston. In order to produce a 50-ns all-atomistic molecular simulation of Trp-cage using AA-REMD a total of 200,000 computing hours (CH) was needed.

All-atomistic molecular simulations with MultiSCAAL consist of three parts. In the first part, the coarse-grained molecular simulation with REM requires ~20,000 CH to complete. In the second part, for the protein reconstruction, the computational burden is negligible compared to the total length of the simulations (< 100CH). In the final part, 1-ns standard all-atomistic molecular dynamics simulation for each reconstructed protein structure as initial conditions takes ~64,000 CH. Upon the completion of MultiSCAAL, 1,280-ns of simulation data is produced in ~84,000 CH. MultiSCAAL simulation provided a considerably enhanced sampling efficiency and lower computational cost than the standard AA-REMD simulations with the total simulation length being ~25 (1280/50) times greater in less computational hours.

## IV. Conclusion

We developed an effective MultiSCAAL method that connects protein models in different resolutions and integrates the potential of mean force from all-atomistic molecular dynamics simulations and energy functions for coarse-grained molecular simulations. Our method can be used to enhance the sampling efficiency of protein conformations in a board range of phase space, particularly in the presence of chemical interference that would be otherwise very difficult to simulate by molecular dynamics alone. Using Trp-cage as a protein model, the folding energy landscape in aqueous condition and at high concentration of urea constructed by MultiSCAAL is broader than AA-REMD, indicating that MultiSCAAL is able to produce a wider distribution of ensemble structures. The NOE distances of the ensemble conformations under high concentration of urea in the dominant structural cluster obtained by MultiSCAAL matched better with NMR experiments than AA-REMD. MultiSCAAL is effective because it uses a reduced representation of side chain beads in a coarse-grained protein model without explicit solvent molecules and this allows it to explore different side-chain orientation much faster than all-atomistic protein models in the presence of excessive water molecules. The algorithm of MultiSCAAL provides genuine transition between high-resolution and low-resolution protein models which helps overcome local entropic traps due to solvation of large side chains in different orientations. In addition, the energy function in the presence of urea for coarse-grained molecular simulations is built from the potential of mean force of all-atomistic molecular dynamics simulations, Boltzmann inversion method and statistical potential and that expedites the computational process needed for equilibrium. Modeling and computation of protein interactions in a cell has been a big challenge, however, MultiSCAAL is a promising and effective sampling scheme we developed for multiscale investigation of chemical interference on protein interactions.

## V. Acknowledgement

MSC would like to thank the National Science Foundation (MCB 0919974) and the University of Houston for the support of this research. AS would like to thank the UH Writing Center for improving the readability of this manuscript. We also thank the Texas Advanced Computing Center and the Texas Learning and Computation Center at UH for providing computational resources.

**Tables:**

**Table 1**. List of the heavy atoms used in SCAAL to represent the side chain of each amino acid.

| Amino acid | Side chain atom | Amino acid | Side chain atom |
|---|---|---|---|
| GLY | ------ | ASN | $C_\gamma$ |
| PHE | $C_\gamma$ | ASP | $C_\beta$ |
| TRP | $C_{\delta 2}$ | THR | $C_\beta$ |
| TYR | $C_\gamma$ | ALA | $C_\beta$ |
| GLN | $C_\gamma$ | GLU | $C_\gamma$ |
| SER | $C_\beta$ | PRO | $C_\beta$ |
| VAL | $C_\beta$ | ARG | $C_\delta$ |
| ILE | $C_{\gamma 1}$ | HIS | $C_\gamma$ |
| LEU | $C_\gamma$ | MET | $S_\delta$ |
| LYS | $C_\delta$ | CYS | $S_\gamma$ |

**Table 2**. 1[st] row in bold: NOE distances reported in reference[24], from photo–CIDNP (Chemically Induced Dynamic Nuclear Polarization) experiments in 6M urea. Rows 2 to 4: Distances calculated between the same protons as in the experiments, for the first, second and third dominant clusters from AA-REMD simulations in 8M urea. Rows 5 to 7: Distances calculated between the same protons as in the experiments, for the first, second and third dominant clusters from MultiSCAAL simulations in 8M urea.

| | Trp6 HE3 Ile4 HB | Trp6 HE3 Ile 4 HG2 | Trp6 HE3 Ile 4 HD1 | Trp6 HE3 Leu 7 HD1 | Trp6 HE3 Leu 7 HD2 | Trp6 HH2 Pro 12 HG2 | Trp6 HE1 Arg 16 HB2 | Trp6 HE1 Arg 16 HB3 |
|---|---|---|---|---|---|---|---|---|
| **NMR (Å)** | **4.2** | **4.5** | **4.5** | **4.0** | **3.4** | **3.3** | **4.2** | **3.8** |
| **AA-REMD (Å) 1[st]** | 8.0 | 8.0 | 8.3 | 3.95 | 8.1 | 4.4 | 7.0 | 7.9 |
| **AA-REMD (Å) 2[nd]** | 8.6 | 9.5 | 8.9 | 3.7 | 7.6 | 4.3 | 7.0 | 6.7 |
| **AA-REMD(Å) 3[d]** | 8.8 | 9.8 | 8.8 | 3.9 | 8.0 | 8.3 | 10.1 | 11.2 |
| **MULTISCAAL (Å) 1[st]** | 6.8 | 6.1 | 6.6 | 4.0 | 6.5 | 4.0 | 3.8 | 3.6 |
| **MULTISCAAL (Å) 2[nd]** | 8.6 | 8.8 | 8.6 | 3.8 | 9.6 | 6.3 | 5.4 | 6.5 |
| **MULTISCAAL (Å) 3[d]** | 7.6 | 8.5 | 9.5 | 4.1 | 9.1 | 8.5 | 7.4 | 6.2 |

**Captions:**

**Figure 1**: A schematic diagram in a multiscale algorithm where a protein configuration switches from all-atomistic (AA) to coarse-grained (CG) representation and vice versa. A side chain-$C_\alpha$ model (SCM) is used as a coarse-grained model. The reconstruction of a protein in an AA representation from CG representation is achieved by SCAAL. The Lennard-Jones (LJ) parameters for an AA representation follow an AMBER force field, while for a CG representation they follow a statistical potential based on bioinformatics and the potential of mean force from the AA molecular dynamic simulations via Boltzmann inversion method. The dynamics of an AA protein is governed by the Newtonian equations of motion. The dynamics of a CG protein is governed by the Langevin/Brownian equations of motion.

**Figure 2:** (a) A schematic representation of the SCAAL reconstruction method with the use of an all-atomistic protein structure as a template and the positions of coarse-grained side-chain-$C_\alpha$ model (SCM) as harmonic constraints. (Left) $C_\alpha$ beads are in red and the heavy side-chain beads are in yellow. The two beads hold the positions through harmonic constraints for a projected reconstructed all-atomistic protein model. A randomly chosen all-atomistic protein structure that can be far from the crystal structure is introduced as a structural template and shown in a solvent accessible surface area mode. (Right) After the structural reconstruction by SCAAL, an all-atomistic representation of a projected protein structure is created (myoglobin, PDBID 1A6M, is used for illustration). (b) A flow chart of the SCAAL reconstruction method.

**Figure 3**: (a) Pair correlation function $g_{ij}^{aq}(r)$ for Thr-Thr (solid line) and Val-Val pairs (dotted line) derived from all–atomistic molecular dynamics simulations in aqueous condition. $r = r^*$

denotes where $g_{ij}^{aq}(r^*)$ is a maximum. (b) The potential of mean force $U_{TT}^{aq}(r)$ of Thr-Thr interaction (in black) is obtained from all–atomistic molecular dynamics simulations under aqueous condition through Boltzmann inversion (eqn 1.). $r$ is the interacting distance between the chosen heavy atoms (i.e. $C_\beta$ atom for Thr. See Table 1) that are in closest proximity to the center of mass of the side chain in threonine. $r^*$ denotes the position of the major peak of the pair correlation function $g^{aq}{}_{TT}(r)$ in (a) and $\varepsilon'^{aq}_{TT} = U_{TT}^{aq}(r^*)$. The Betancourt-Thirumalai statistical potential follows a Lennard-Jones interaction $V_{TT}^{aq}(r)$ (eqn. 3) for the same pair of amino acid in coarse-grained molecular simulations (in red). $r$ is the interacting distance between the coarse-grained side chain beads of the amino acids (i.e. center of mass of side chains). $r_o$ is the bonding distance $\sigma_{TT}$ in eqn S1. $\varepsilon_{TT}^{aq} = V_{TT}^{aq}(r_o)$ is taken from the Betancourt–Thirumalai statistical potential. The reference potential from eqn 2 is $V_o$. (c) $V_{ij}^{aq}(r)$ for Thr-Thr (solid line) and Val-Val pairs (dotted line) in aqueous solvent. $r_o$ is the same bonding distance in (b).

**Figure 4:** The correlation between the aqueous solvent-mediated interactions between amino acids $i$ and $j$, $\varepsilon'^{aq}_{ij}$, which are derived from the molecular dynamics simulations and the ones from the Betancourt-Thirumalai statistical potential $\varepsilon_{ij}^{aq}$. The linear correlation coefficient is 0.79.

**Figure 5**: Two–dimensional free energy landscape of Trp-cage as a function of the radius of gyration ($R_g$) and the root-mean-square deviation (RMSD) under (a) aqueous and (b) urea conditions based on coarse-grained molecular simulations at 300K. The free energy is colored by $k_BT$ and the color of an area with values greater than 11 is white.

**Figure 6:** Representative structures from the dominant clusters calculated under (a, c, e) aqueous and (b, d, f) urea conditions based on three different simulation schemes: (a, b) Coarse-grained molecular simulations; (c, d) All-atomistic molecular dynamics simulations with replica exchange method (AA-REMD); (e, f) All-atomistic molecular dynamics simulations using the Multi-SCAAL. Residues Tyr3, Trp6, and Leu7 are shown in yellow. Residues Pro12, Pro17, Pro18, and Pro19 are shown in silver. The trace of the protein backbone is colored from red (N-terminus) to blue (C-terminus).

**Figure 7**: The average root-mean-square deviation (RMSD) of heavy atoms in a residue between the original and reconstructed unfolded protein structures using SCAAL (red) and PULCHRA (blue).

**Figure 8**: Two–dimensional free energy landscape for Trp-cage as a function of the radius of gyration ($R_g$) and the root-mean-square-deviation (RMSD) under (a, c) aqueous and (b, d) urea conditions based on two different simulation schemes at 300K: (a, b) simulations using AA-REMD; (c, d) simulations using MultiSCAAL. The free energy is colored by $k_BT$.

**Figure 9:** The first, second, and third most dominant structures of Trp-cage under aqueous conditions clustered from (a, b, c) all-atomistic molecular dynamics simulations with the replica exchange method (AA-REMD) and (d, e, f) the all-atomistic molecular simulation using MultiSCAAL. The percentages of conformations in each cluster are included. The dominant structures in (a, d) are the same as the ones shown in Fig. 6 (c, e), respectively, for illustration purposes.

**Figure 10:** The first, second, and third dominant structures of Trp-cage under 8M urea conditions clustered from (a, b, c) all-atomistic molecular dynamics simulations with the replica exchange method (AA-REMD) and (d, e, f) MultiSCAAL. The percentages of conformations in each cluster are  included. The dominant structures in (a, d) are the same as the ones shown in Fig. 6 (d, f), respectively, for illustration purposes.

**Figure 11:**  A representative structure of Trp–cage under 8M urea from the most dominant cluster from MultiSCAAL simulations. Residues Ile4 and Trp6 are shown in yellow, while protons HE3 in Trp6 and HG2 in Ile4 are colored silver. Water molecules located in the hydrophobic core are shown in van der Waals representation.

Figures

Fig 1

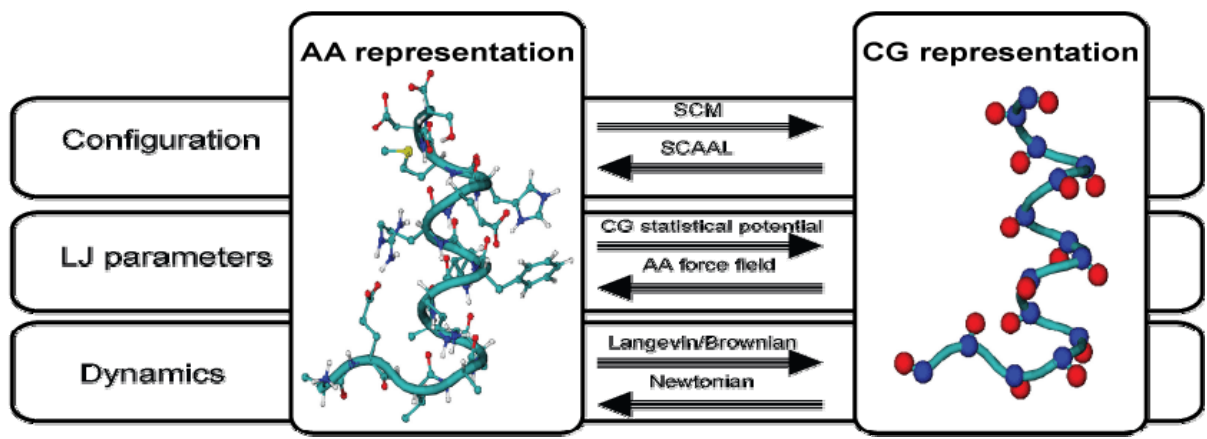

Fig 2

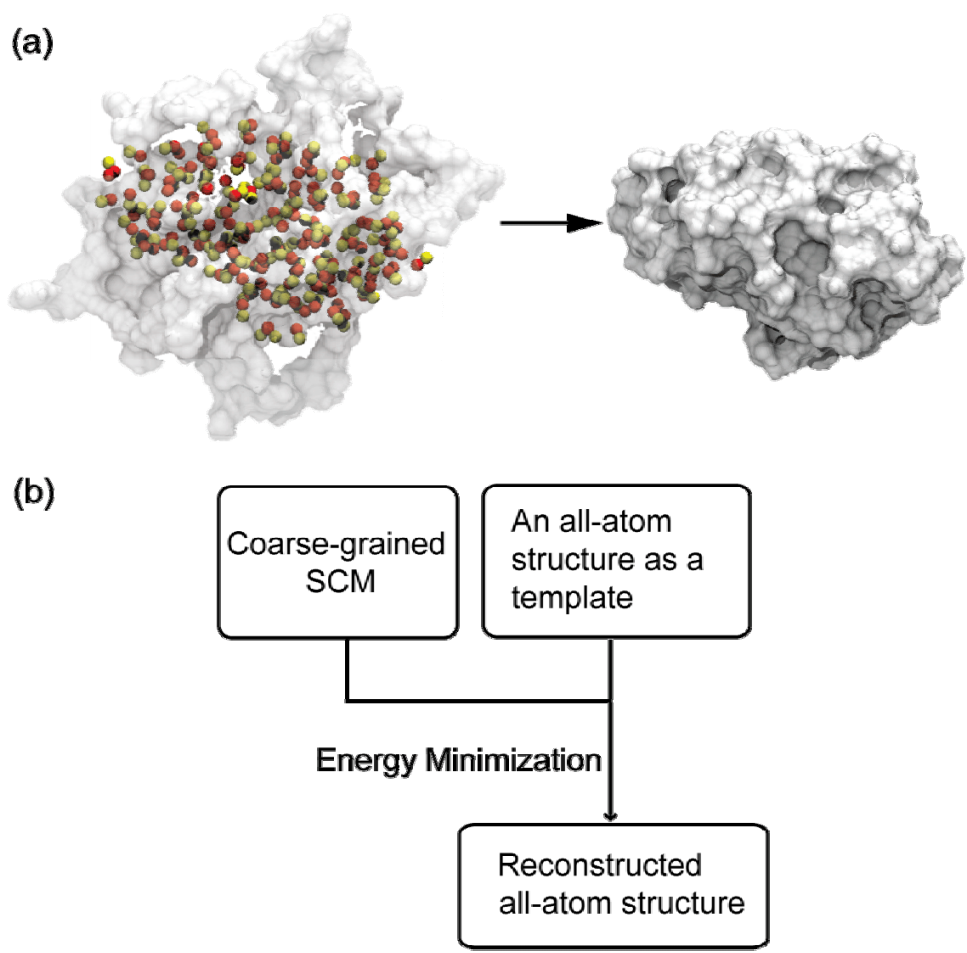

Fig 3

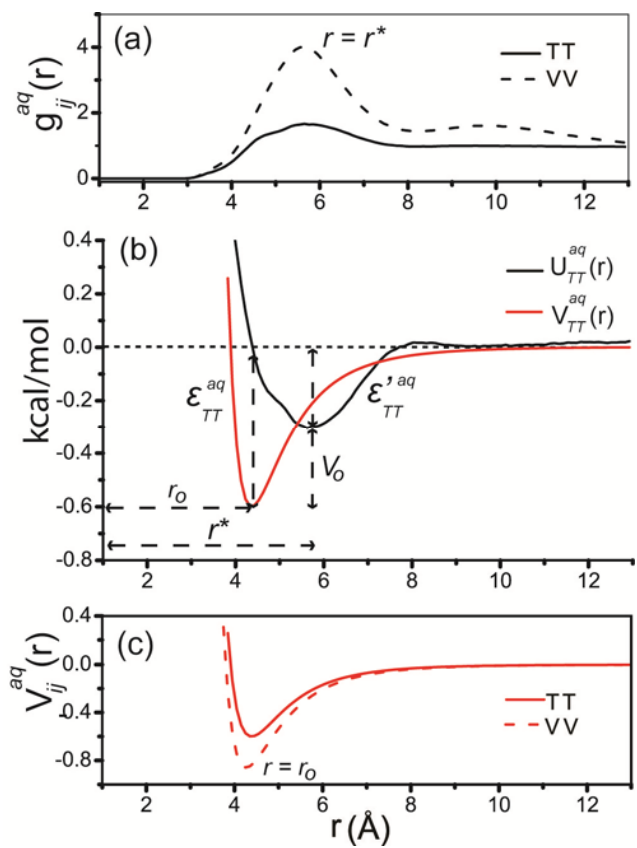

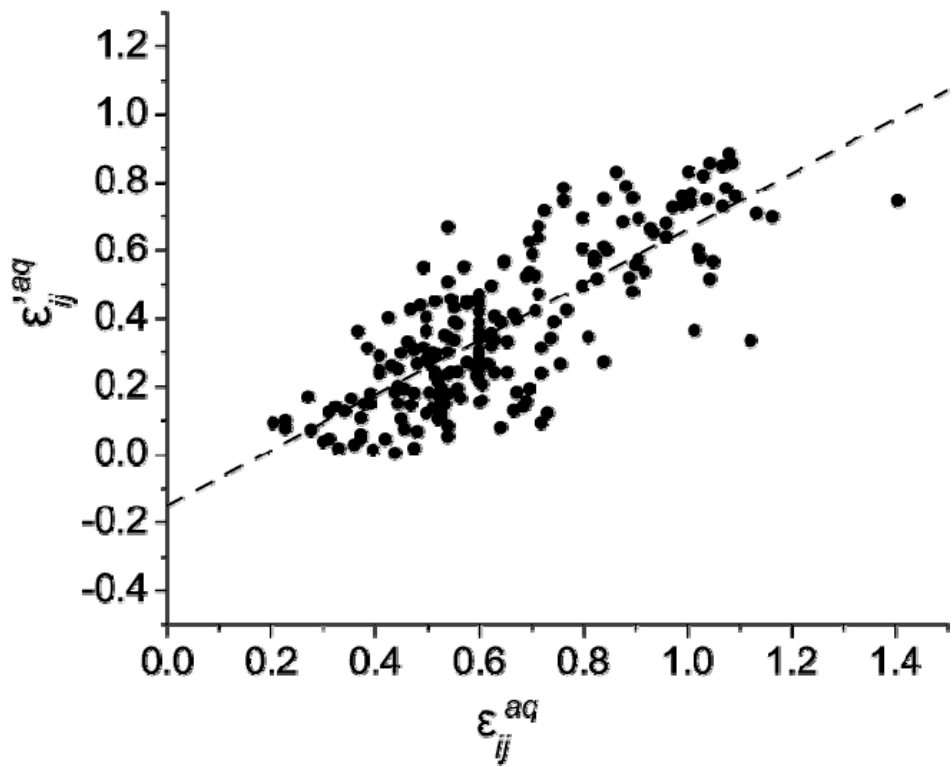

Fig 5

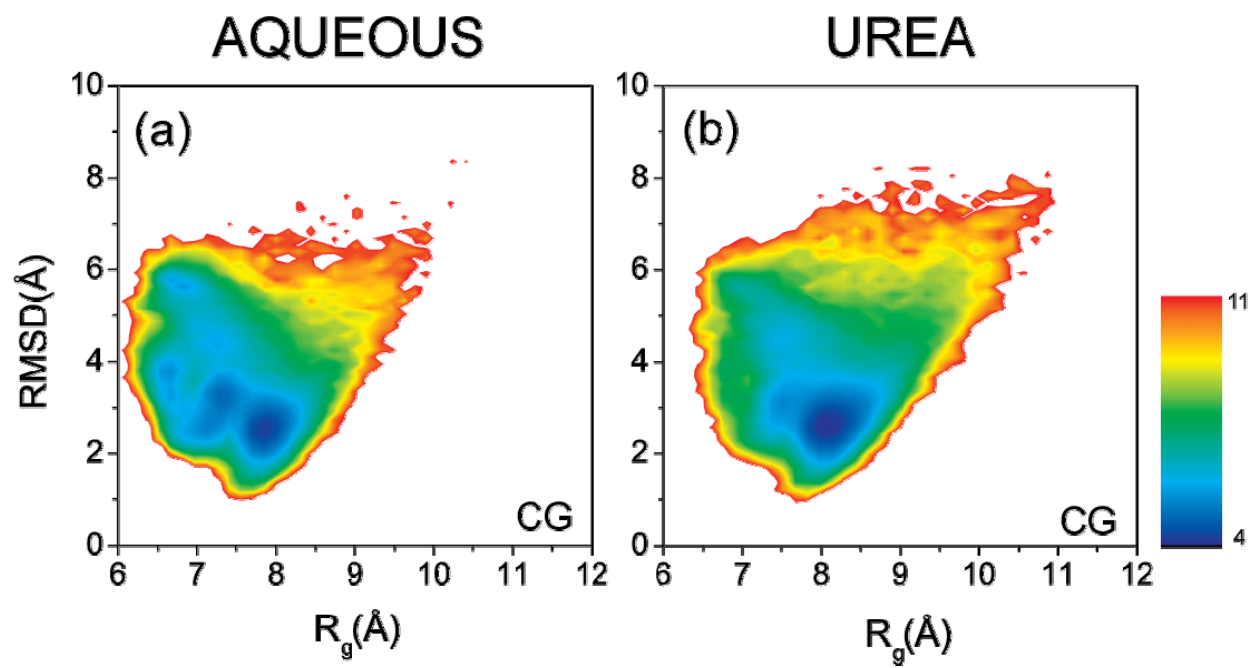

Fig 6

AQUEOUS                    UREA

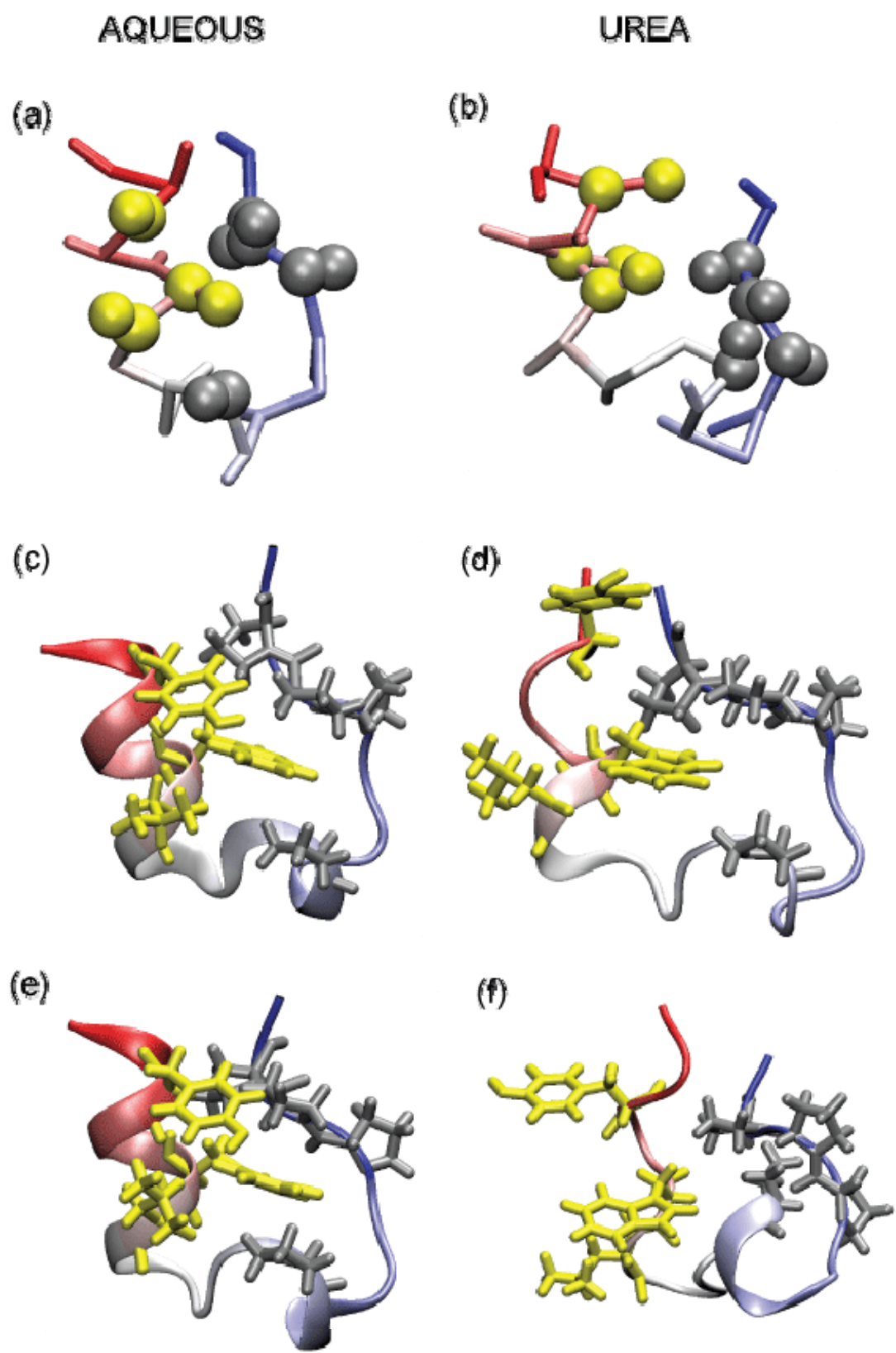

Fig 7

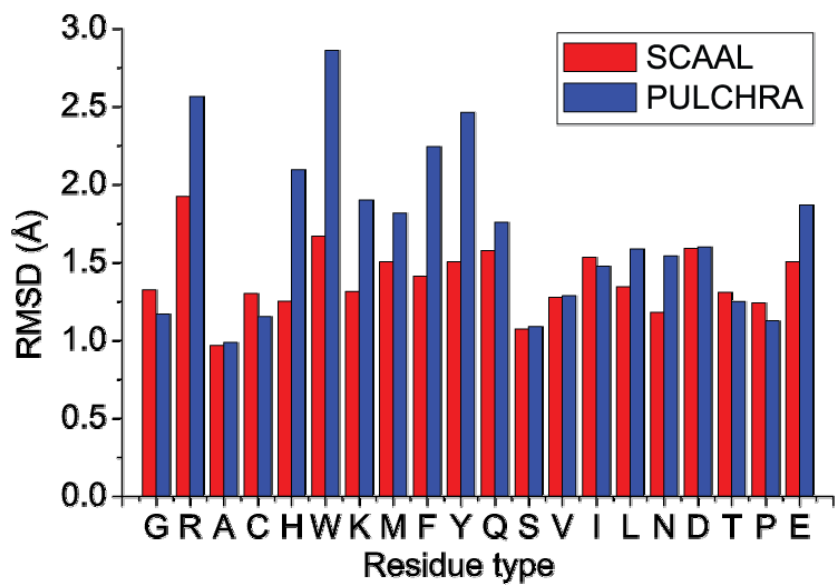

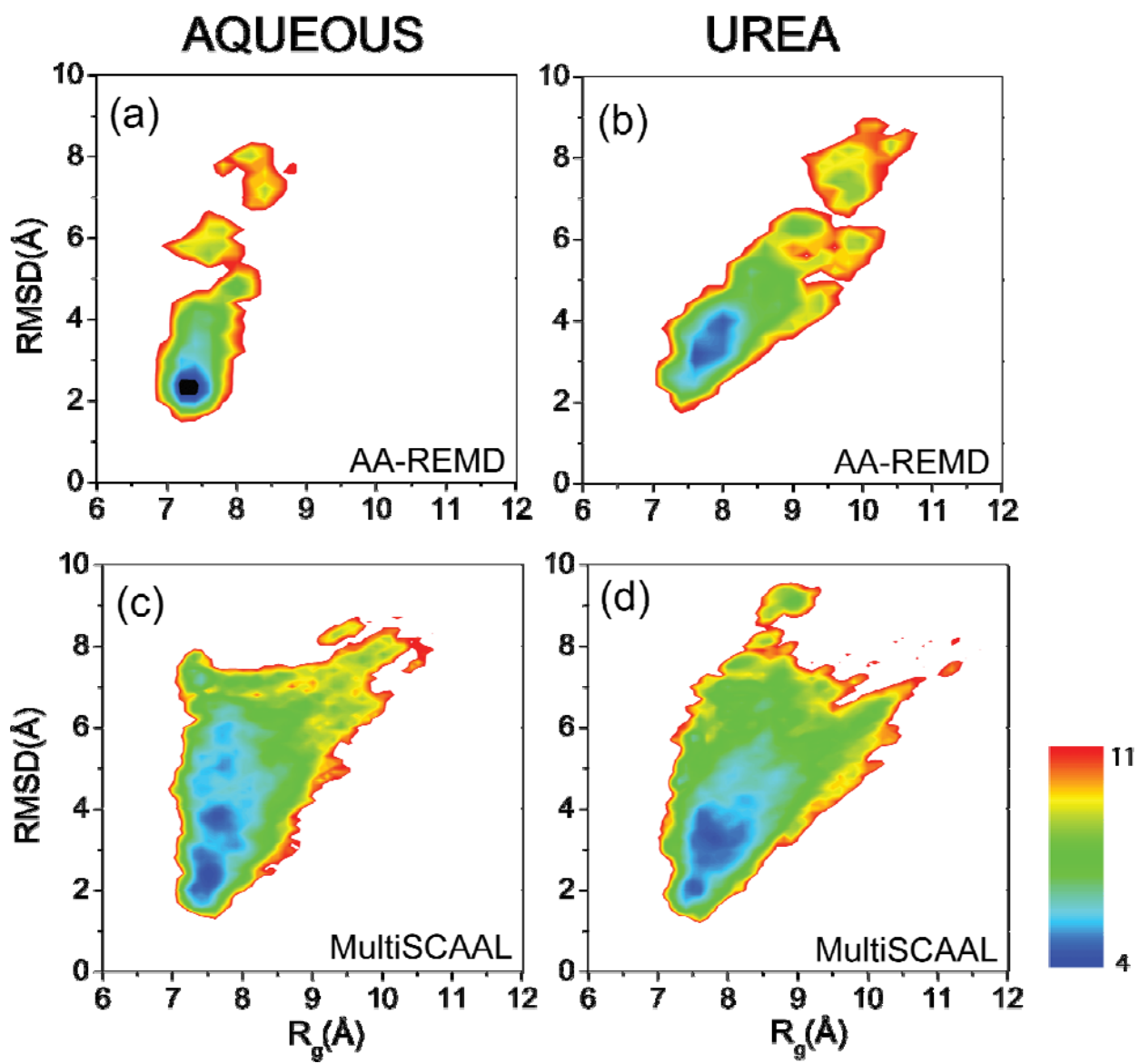

Fig 9

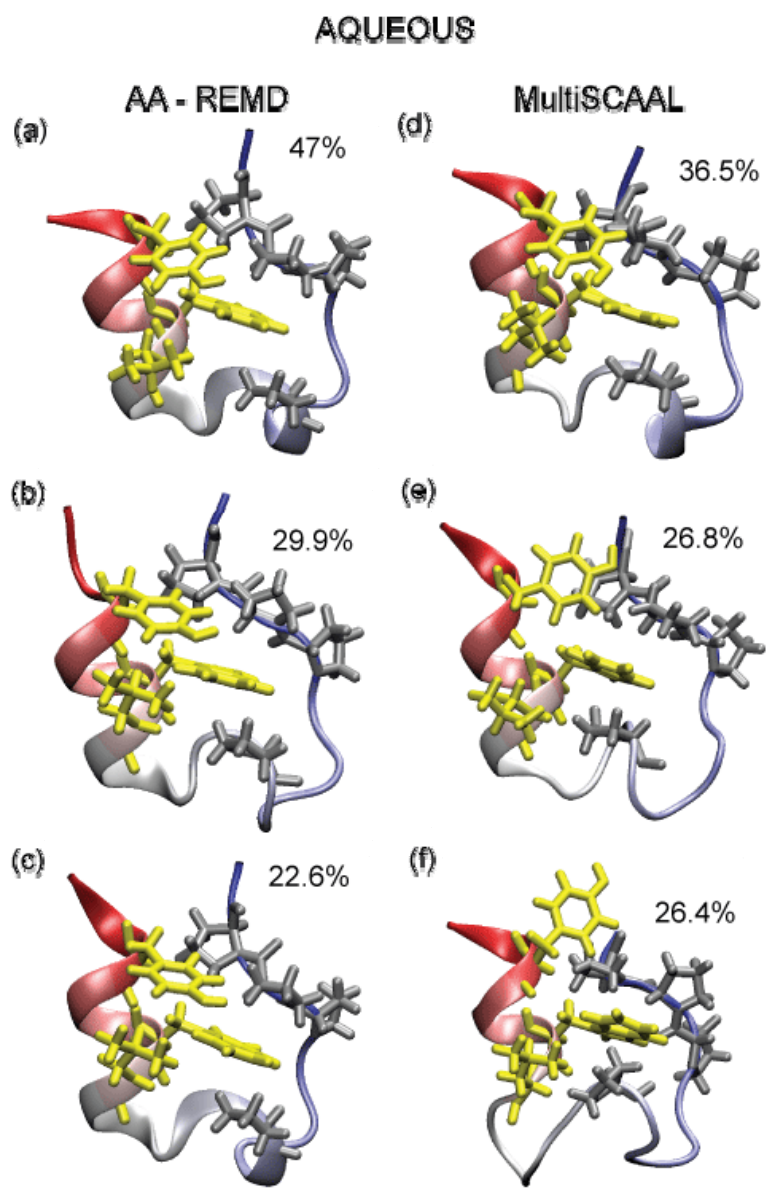

Fig 10

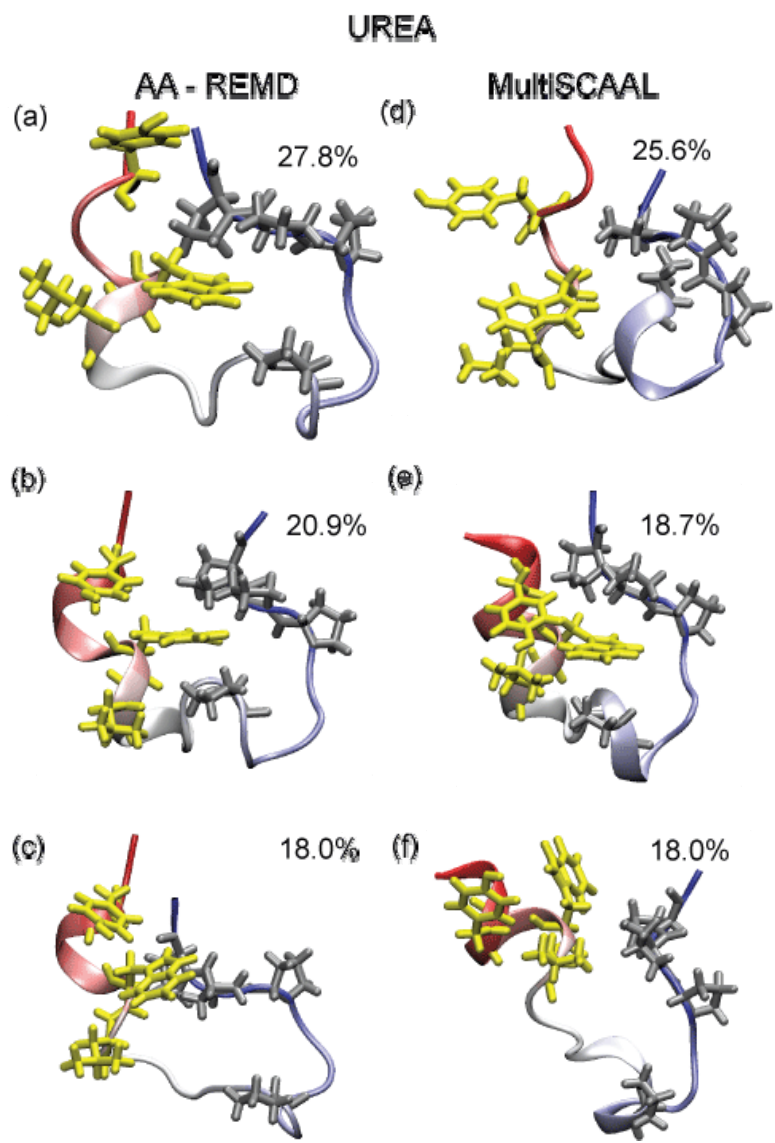

UREA

(a) AA - REMD     (d) MultISCAAL

(a) 27.8%     (d) 25.6%

(b) 20.9%     (e) 18.7%

(c) 18.0%     (f) 18.0%

Fig 11

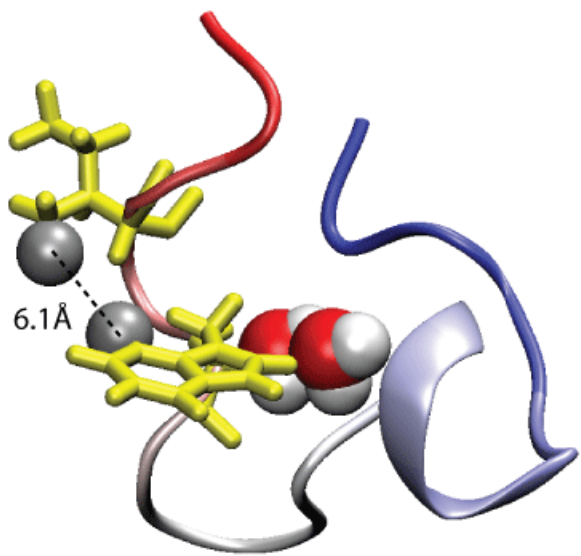

REFERENCE

[1]   R. J. Ellis, Curr. Opin. Struct. Biol. **11** (2001).
[2]   L. Hua, R. H. Zhou, D. Thirumalai, and B. J. Berne, Proc Natl Acad Sci USA **105**, 16928 (2008).
[3]   H. Kokubo, J. Rosgen, D. W. Bolen, and B. M. Pettitt, Biophysical Journal **93**, 3392 (2007).
[4]   R. Zangi, R. H. Zhou, and B. J. Berne, J. Am. Chem. Soc. **131**, 1535 (2009).
[5]   A. M. Fernandez-Escamilla, M. S. Cheung, M. C. Vega, M. Wilmanns, J. N. Onuchic, and S. L., Proc Natl Acad Sci USA **101**, 2834 (2004).
[6]   D. Homouz, M. Perham, A. Samiotakis, M. S. Cheung, and P. Wittung-Stafshede, Proc Natl Acad Sci USA **105**, 11754 (2008).
[7]   D. Homouz, L. Stagg, P. Wittung-Stafshede, and M. S. Cheung, Biophysical Journal **96**, 671 (2009).
[8]   L. Stagg, S.-Q. Zhang, P. Wittung-Stafshede, and M. S. Cheung, Proc Natl Acad Sci USA **104**, 18976 (2007).
[9]   C. D. Snow, H. Nguyen, V. S. Pande, and M. Gruebele, Nature **420**, 102 (2002).
[10]  J. N. Onuchic and P. G. Wolynes, Curr. Opin. Struct. Biol. **14** (2004).
[11]  M. S. Cheung, L. L. Chavez, and J. N. Onuchic, Polymer **45** (2004).
[12]  W. C. Still, A. Tempczyk, R. C. Hawley, and T. Hendrickson, J. Am. Chem. Soc. **112**, 6127 (1990).
[13]  J. Chen, C. L. Brooks, and J. Khandogin, Curr. Opin. Struct. Biol. **18**, 140 (2008).
[14]  S. Izvekov and G. A. Voth, J. Phys. Chem. B **109**, 2469 (2005).
[15]  W. G. Noid, J.-W. Chu, G. S. Ayton, V. Krishna, S. Izvekov, G. A. Voth, A. Das, and H. C. Andersen, J. Chem. Phys. **128**, 244114 (2008).
[16]  W. G. Noid, P. Liu, Y. Wang, J.-W. Chu, G. S. Ayton, S. Izvekov, H. C. Andersen, and G. A. Voth, J. Chem. Phys. **128**, 244115 (2008).
[17]  S. Matysiak, C. Clementi, M. Praprotnik, K. Kremer, and L. Delle Site, J. Chem. Phys. **128** (2008).
[18]  M. Praprotnik, L. Delle Site, and K. Kremer, Phys Rev E **73** (2006).
[19]  B. Ensing, S. O. NIelsen, P. B. Moore, M. L. Klein, and M. Parrinello, J. Chem. Theory Comput. **3**, 11001105 (2007).
[20]  M. Neri, C. Anselmi, M. Cascella, A. Maritan, and P. Carloni, Phy Rev Lett **95**, 218102 (2005).
[21]  Y. Zhang, M. H. Peters, and Y. Li, Proteins: Struct. Funct. Genet. **52**, 339 (2003).
[22]  J. Neidigh, W,, R. M. Fesinmeyer, and N. H. Andersen, Nat Struct Biol **9**, 425 (2002).
[23]  W. W. Streicher and G. I. Makhatadze, Biochemistry **46**, 2876 (2007).
[24]  K. H. Mok, L. T. Kuhn, M. Goez, I. Day, L. J., N. H. Andersen, and P. J. Hore, Nature **447**, 106 (2007).
[25]  Z. Gattin, S. Riniker, P. J. Hore, K. H. Mok, and W. F. van Gunsteren, Protein Science **18**, 2090 (2009).
[26]  D. Paschek, H. Nymeyer, and A. E. Garcia, J. Struct. Biol. **157**, 524 (2007).
[27]  D. Paschek, S. Hempel, and A. E. Garcia, Proc Natl Acad Sci USA **46**, 17754 (2008).
[28]  R. Zhou, Proc Natl Acad Sci USA **23**, 13280 (2003).
[29]  C. Simmerling, B. Strockbine, and A. Roitberg, J. Am. Chem. Soc. **124**, 11258 (2002).
[30]  C. D. Snow, B. Zagrovic, and V. S. Pande, J. Am. Chem. Soc. **124** (2002).

[31]  B. Zagrovic and V. S. Pande, J Comput Chem **24**, 1432 (2003).
[32]  M. S. Cheung, J. M. Finke, B. Callahan, and J. N. Onuchic, J. Phys. Chem. B **107**, 11193 (2003).
[33]  D. Klimov and D. Thirumalai, Proc Natl Acad Sci USA **97**,  2544 (2000).
[34]  V. Daggett, Acc. Chem. Res. **35**, 422 (2002).
[35]  H. A. Scheraga, M. Khalili, and A. Liwo, Curr. Opin. Struct. Biol. **58**, 57 (2007).
[36]  M. R. Betancourt and S. J. Omovie, J. Chem. Phys. **130**, 195103 (2009).
[37]  D. Reith, M. Putz, and F. Muller-Plathe, J Comput Chem **24**, 1624 (2003).
[38]  D. Frenkel and B. Smit, *Understanding Molecular Simulation: From Algorithms to Applications* (Academic Press, 2001).
[39]  W. Li and S. Takada, J. Chem. Phys. **130** (2009).
[40]  Z. Z. Fan, J.-K. Hwang, and A. Warshel, Theor. Chem. Acc. **103**, 77 (1999).
[41]  M. R. Betancourt and D. Thirumalai, Protein Science **8**, 361 (1999).
[42]  M. J. Sippl, J. Mol. Biol. **213** (1990).
[43]  A. K. Sope, Chem. Phys. **202**, 295 (1996).
[44]  M. Makowski, A. Liwo, M. K., J. Makowska, and H. A. Scheraga, J. Phys. Chem. B **111**, 2917 (2007).
[45]  D. A. Case, et al., University of California, San Fransisco  (2006).
[46]  W. L. Jorgensen, J. Chandrasekhar, J. Madura, R. W. Impey, and M. L. Klein, J. Chem. Phys. **79**, 926 (1983).
[47]  T. Darden, D. York, and L. Pedersen, J. Chem. Phys. **98**, 10089 (1993).
[48]  H. C. Andersen, J. Chem. Phys. **72**, 2384 (1980).
[49]  T. Veitshans, D. Klimov, and D. Thirumalai, Fold. Des. **2**, 1 (1997).
[50]  J. L. Fauchere and V. Pliska, Eur. J. Med. Chem. **18**, 369 (1983).
[51]  Y. Sugita and Y. Okamoto, Chem. Phys. Lett. **314**, 141 (1999).
[52]  D. Case, et al., *AMBER6* (UCSF, 1999).
[53]  M. E. J. Newman and G. T. Barkema, *Monte Carlo Methods in Statistical Physics* (Oxford University Press, 1999).
[54]  M. Feig, P. Rotkiewicz, A. Kolinski, J. Skolnick, and C. L. Brooks, Proteins: Struct. Funct. Genet. **41**, 86 (2000).
[55]  E. Lyman, F. M. Ytreberg, and D. M. Zuckerman, Phy Rev Lett **96** (2006).
[56]  E. Lyman and D. M. Zuckerman, J Chem Theory Comp. **2**, 656 (2006).
[57]  V. Hornak, R. Abel, A. Okur, B. R. Strockbine, A., and C. Simmerling, Proteins: Struct. Funct. Bioinf. **65**, 712 (2006).
[58]  M. Miyazawa and R. L. Jernigan, J. Mol. Biol. **256**, 623 (1996).
[59]  M. E. Karpen, D. J. Tobins, and C. L. Brooks, Biochemistry **32**, 412 (1993).
[60]  G. A. Carpenter and S. Grossberg, Appl. Opt. **26**, 4919 (1987).
[61]  Z. Guo and D. Thirumalai, Fold. Des. **2**, 377 (1997).
[62]  M. S. Cheung and D. Thirumalai, J. Phys. Chem. B **111** (2007).
[63]  P. Rotkiewicz and J. Skolnick, J Comput Chem **29**, 1460 (2008).
[64]  A. Canutescu, A. Shelenkov, and R. Dunbrack, Protein Science **12**, 2001 (2003).
[65]  A. P. Heath, L. E. Kavraki, and C. Clementi, Proteins: Struct. Funct. Bioinf. **68**, 646 (2007).

Multiscale Investigation of Chemical Interference in Protein Dynamics

Antonios Samiotakis*[+], Dirar Homouz*[+] and Margaret S. Cheung*

METHODS : *Energy Function for Coarse-Grained Protein Model:*

A Sidechain-$C_\alpha$ (SCM)[1] coarse-grained model is used to represent proteins where each amino acid (except glycine) is modeled by two beads: a $C_\alpha$ bead and a side chain bead. The potential energy of a protein, $E_p$ is $E_{structural}+E_{HB}+E_{NB}$ where the structural energy, $E_{structural}$, is the sum of bond-length potential, side chain-backbone connectivity potential, bond-angle potential, dihedral potential, and chiral interactions. The chiral energy accounts for an L-isoform preference of side chains. Each term is described in previous studies[2-4].

Nonbonded interactions $E^{NB}_{ij}$ between a pair of i and j side chain beads at a distance *r* are as follows,

$$E_{ij}^{NB} = \varepsilon_{ij}[(\frac{\sigma_{ij}}{r_{ij}})^{12} - 2(\frac{\sigma_{ij}}{r_{ij}})^{6}]$$

(S1)

where $\sigma_{ij} = f(\sigma_i + \sigma_j)$. $\sigma_i$ and $\sigma_j$ are the Van der Waals (VdW) radii of side chain beads. To avoid volume clash, f = 0.9 and |i-j|>2. $\varepsilon_{ij}$, is based on the solvent-mediated interaction between pairs of residues (see below).

For backbone hydrogen bonding interactions, an angular-dependent function that captures directional properties of backbone hydrogen bonds is used in:

$$E_{ij}^{HB} = A(\rho)E_{ij}^{NB}$$

(S2)

$$A(\rho) = \frac{1}{[1 + (1 - \cos^2 \rho)(1 - \frac{\cos \rho}{\cos \rho_a})]^2} \qquad (S3)$$

where $E^{NB}_{ij}$ shares the same formula as eqn (S1), except that $\varepsilon_{ij}$ for backbone hydrogen bonding is 0.6 kcal/mol and $\sigma_{ij}$ is the hydrogen bond length, 4.6 Å .

$A(\rho)$ in eqn S3. measures the structural alignment of two interacting strands. $\rho$ is the pseudo dihedral angle between two interacting strands of backbones[1]. $A(\rho)=1$ if the alignment points to β-strands or α-helices. $\rho_a$ is the pseudo dihedral angle of a canonical helical turn, 0.466 (rad).

*The equations of motion for the coarse-grained protein model:*

To account for the effect of the solvent on the protein dynamics we use the Langevin equation of motion[5] (Eq. (S4)) to describe the dynamics in our coarse-grained molecular simulations. The solvent is treated implicitly in the Langevin equation through a stochastic term. The Langevin equation of motion for a general coordinate x is:

$$m\ddot{x} = -\frac{\partial U}{\partial x} - \zeta \dot{x} + \Gamma , \qquad (S4)$$

where m is the mass and U is the potential energy of the molecule. The drag term, $-\zeta \dot{x}$, or the dissipation term, is caused by friction which is compensated by a random force Γ representing random collisions with solvent molecules. Γ is taken from a distribution of a white noise (Gaussian noise).

Fast motions in large biomolecules are quickly damped in a viscous solvent such as water. As a result, they follow random trajectories referred to as the Brownian motion. The inertia term is dropped in Eq. (S4) and we get the first order ordinary differential equation for the Brownian motion in eqn S5:

$$\zeta \dot{x} = -\frac{\partial U}{\partial x} + \Gamma \ . \tag{5}$$

*Protein structures used to test SCAAL*

The 67 non-redundant structures (SCOP characterizations of all-alpha, all − beta, alpha/beta and alpha+beta proteins) with less than 30% sequence similarity used in testing the accuracy and overall performance of SCAAL are: 1A6M 1BYI 1C75 1C7K 1CEX 1DY5 1EB6 1F9Y 1G2Y 1G66 1GA6 1GCI 1GKM 1I1W 1IQZ 1IX9 1IXH 1J0P 1JFB 1K4I 1K5C 1KWF 1L9L 1LNI 1LUG 1M1Q 1MJ5 1MN8 1MUW 1MWQ 1N4W 1N55 1NKI 1NLS 1NQJ 1NWZ 1O7J 1OAI 1OD3 1OEW 1OK0 1P1X 1PJX 1PQ7 1Q6Z 1R2M 1R6J 1RTQ 1SFD 1SSX 1TQG 1TT8 1UCS 1UFY 1UG6 1UNQ 1US0 1UZV 1VYR 1W0N 1X8Q 1XG0 2ERL 2FDN 2PVB 3LZT and 7A3H.

TABLES:

|     | Cys | Phe | Leu | Trp | Val | Ile | Met | His | Tyr | Ala | Gly | Pro | Asn | Thr | Ser | Arg | Gln | Asp | Lys | Glu |
|-----|-----|-----|-----|-----|-----|-----|-----|-----|-----|-----|-----|-----|-----|-----|-----|-----|-----|-----|-----|-----|
| Cys | 0.60 |     |     |     |     |     |     |     |     |     |     |     |     |     |     |     |     |     |     |     |
| Phe | 0.65 | 0.77 |     |     |     |     |     |     |     |     |     |     |     |     |     |     |     |     |     |     |
| Leu | 0.61 | 0.82 | 0.91 |     |     |     |     |     |     |     |     |     |     |     |     |     |     |     |     |     |
| Trp | 0.57 | 0.57 | 0.54 | 0.44 |     |     |     |     |     |     |     |     |     |     |     |     |     |     |     |     |
| Val | 0.66 | 0.75 | 0.90 | 0.65 | 0.88 |     |     |     |     |     |     |     |     |     |     |     |     |     |     |     |
| Ile | 0.54 | 0.68 | 0.82 | 0.71 | 0.79 | 0.77 |     |     |     |     |     |     |     |     |     |     |     |     |     |     |
| Met | 0.60 | 0.68 | 0.75 | 0.70 | 0.72 | 0.59 | 0.63 |     |     |     |     |     |     |     |     |     |     |     |     |     |
| His | 0.57 | 0.63 | 0.72 | 0.71 | 0.68 | 0.55 | 0.55 | 0.66 |     |     |     |     |     |     |     |     |     |     |     |     |
| Tyr | 0.61 | 1.07 | 0.79 | 0.64 | 0.72 | 0.69 | 0.64 | 0.62 | 0.68 |     |     |     |     |     |     |     |     |     |     |     |
| Ala | 0.52 | 0.62 | 0.70 | 0.61 | 0.68 | 0.55 | 0.53 | 0.52 | 0.53 | 0.52 |     |     |     |     |     |     |     |     |     |     |
| Gly | 0.54 | 0.46 | 0.64 | 0.47 | 0.65 | 0.50 | 0.54 | 0.54 | 0.54 | 0.54 | 0.54 |     |     |     |     |     |     |     |     |     |
| Pro | 0.58 | 0.66 | 0.65 | 0.69 | 0.70 | 0.63 | 0.66 | 0.72 | 0.48 | 0.44 | 0.53 | 0.54 |     |     |     |     |     |     |     |     |
| Asn | 0.45 | 0.41 | 0.45 | 0.56 | 0.50 | 0.35 | 0.43 | 0.52 | 0.53 | 0.38 | 0.50 | 0.47 | 0.56 |     |     |     |     |     |     |     |
| Thr | 0.52 | 0.55 | 0.63 | 0.44 | 0.61 | 0.55 | 0.53 | 0.56 | 0.52 | 0.50 | 0.50 | 0.49 | 0.45 | 0.49 |     |     |     |     |     |     |
| Ser | 0.50 | 0.43 | 0.47 | 0.55 | 0.50 | 0.43 | 0.48 | 0.53 | 0.43 | 0.39 | 0.48 | 0.45 | 0.49 | 0.45 | 0.37 |     |     |     |     |     |
| Arg | 0.43 | 0.70 | 0.48 | 0.57 | 0.50 | 0.43 | 0.35 | 0.41 | 0.44 | 0.37 | 0.40 | 0.40 | 0.40 | 0.41 | 0.32 | 0.30 |     |     |     |     |
| Gln | 0.48 | 0.56 | 0.52 | 0.55 | 0.51 | 0.45 | 0.42 | 0.48 | 0.49 | 0.41 | 0.42 | 0.48 | 0.46 | 0.49 | 0.43 | 0.39 | 0.38 |     |     |     |
| Asp | 0.36 | 0.50 | 0.23 | 0.26 | 0.32 | 0.35 | 0.35 | 0.34 | 0.32 | 0.26 | 0.31 | 0.31 | 0.36 | 0.46 | 0.50 | 0.87 | 0.35 | 0.30 |     |     |
| Lys | 0.35 | 0.23 | 0.31 | 0.43 | 0.37 | 0.25 | 0.18 | 0.31 | 0.40 | 0.23 | 0.35 | 0.38 | 0.35 | 0.32 | 0.42 | 0.35 | 0.31 | 0.62 | 0.37 |     |
| Glu | 0.38 | 0.37 | 0.37 | 0.31 | 0.34 | 0.18 | 0.36 | 0.36 | 0.36 | 0.30 | 0.33 | 0.37 | 0.32 | 0.45 | 0.49 | 0.77 | 0.38 | 0.30 | 0.59 | 0.30 |

**Table S1**. Statistical potential parameter table (SPPT) representing the Lennard–Jones solvent mediated interactions $\varepsilon_{ij}^{ur}$ in kcal/mol for all the amino acid pairs derived from all–atomistic molecular dynamics simulations in 8M urea solvent.

| | Cys | Phe | Leu | Trp | Val | Ile | Met | His | Tyr | Ala | Gly | Pro | Asn | Thr | Ser | Arg | Gln | Asp | Lys | Glu |
|---|---|---|---|---|---|---|---|---|---|---|---|---|---|---|---|---|---|---|---|---|
| Cys | 1.40 | | | | | | | | | | | | | | | | | | | |
| Phe | 0.92 | 1.09 | | | | | | | | | | | | | | | | | | |
| Leu | 0.90 | 1.07 | 1.09 | | | | | | | | | | | | | | | | | |
| Trp | 1.04 | 1.07 | 1.02 | 1.04 | | | | | | | | | | | | | | | | |
| Val | 0.91 | 1.00 | 1.08 | 0.97 | 1.03 | | | | | | | | | | | | | | | |
| Ile | 0.89 | 0.99 | 1.07 | 0.99 | 1.01 | 0.96 | | | | | | | | | | | | | | |
| Met | 0.89 | 1.13 | 1.01 | 1.16 | 0.88 | 0.96 | 0.94 | | | | | | | | | | | | | |
| His | 0.71 | 0.71 | 0.54 | 0.88 | 0.49 | 0.49 | 0.70 | 0.80 | | | | | | | | | | | | |
| Tyr | 0.70 | 0.89 | 0.86 | 0.93 | 0.76 | 0.80 | 0.91 | 0.73 | 0.76 | | | | | | | | | | | |
| Ala | 0.76 | 0.80 | 0.82 | 0.84 | 0.83 | 0.81 | 0.74 | 0.47 | 0.69 | 0.72 | | | | | | | | | | |
| Gly | 0.65 | 0.53 | 0.52 | 0.74 | 0.58 | 0.47 | 0.55 | 0.46 | 0.62 | 0.62 | 0.72 | | | | | | | | | |
| Pro | 0.71 | 0.71 | 0.65 | 1.04 | 0.65 | 0.57 | 0.70 | 0.63 | 0.84 | 0.56 | 0.61 | 0.64 | | | | | | | | |
| Asn | 0.43 | 0.43 | 0.38 | 0.65 | 0.37 | 0.27 | 0.41 | 0.54 | 0.59 | 0.46 | 0.54 | 0.52 | 0.62 | | | | | | | |
| Thr | 0.60 | 0.60 | 0.60 | 0.60 | 0.60 | 0.60 | 0.60 | 0.60 | 0.60 | 0.60 | 0.60 | 0.60 | 0.60 | 0.60 | | | | | | |
| Ser | 0.55 | 0.54 | 0.44 | 0.56 | 0.45 | 0.39 | 0.41 | 0.51 | 0.56 | 0.51 | 0.54 | 0.50 | 0.52 | 0.60 | 0.52 | | | | | |
| Arg | 0.41 | 0.55 | 0.55 | 0.85 | 0.50 | 0.49 | 0.50 | 0.58 | 0.82 | 0.44 | 0.52 | 0.61 | 0.59 | 0.60 | 0.53 | 0.52 | | | | |
| Gln | 0.58 | 0.62 | 0.55 | 0.67 | 0.50 | 0.52 | 0.61 | 0.47 | 0.71 | 0.47 | 0.48 | 0.63 | 0.63 | 0.60 | 0.45 | 0.67 | 0.52 | | | |
| Asp | 0.37 | 0.31 | 0.23 | 0.56 | 0.20 | 0.28 | 0.23 | 0.73 | 0.64 | 0.42 | 0.50 | 0.45 | 0.67 | 0.60 | 0.59 | 1.03 | 0.53 | 0.44 | | |
| Lys | 0.39 | 0.53 | 0.50 | 0.77 | 0.50 | 0.47 | 0.47 | 0.44 | 0.84 | 0.48 | 0.53 | 0.53 | 0.68 | 0.60 | 0.54 | 0.30 | 0.72 | 1.01 | 0.37 | |
| Glu | 0.32 | 0.40 | 0.38 | 0.69 | 0.35 | 0.37 | 0.46 | 0.67 | 0.70 | 0.34 | 0.31 | 0.44 | 0.61 | 0.60 | 0.54 | 1.05 | 0.54 | 0.36 | 1.12 | 0.33 |

**Table S2**. Interaction energies $\varepsilon_{ij}^{aq}$ expressed in kcal/mol for all the amino acid pairs[6] in aqueous solvent.

CAPTIONS:

**Figure S1.** Normalized probability distribution of the interaction energies from the Betancourt-Thirumalai[6] statistical potential for amino acid pairs $i$ and $j$, $\varepsilon_{ij}^{aq}$ (red) and the solvent-mediated interactions obtained from the potential of mean force from all − atomistic molecular dynamics simulations on amino acid pairs in aqueous solvent, $\varepsilon_{ij}^{'aq}$ (blue).

FIGURES

Fig S1

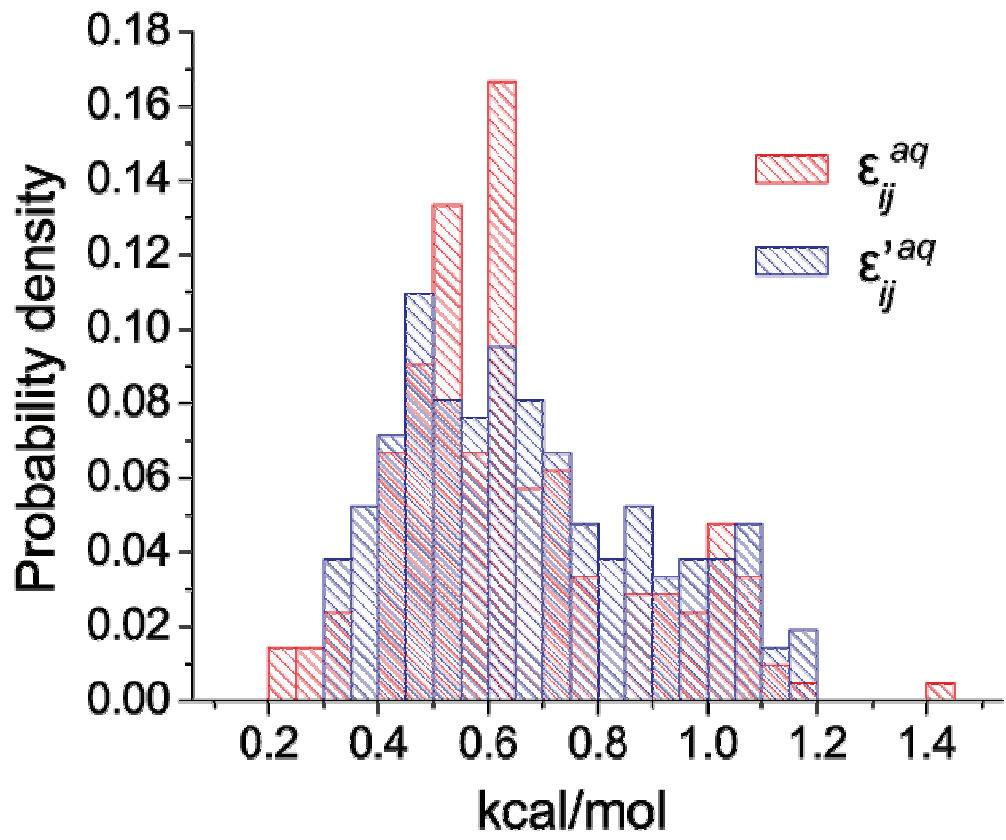

[1]    M. S. Cheung, J. M. Finke, B. Callahan, and J. N. Onuchic, J. Phys. Chem. B **107**, 11193 (2003).

[2]    L. Stagg, S.-Q. Zhang, P. Wittung-Stafshede, and M. S. Cheung, Proc Natl Acad Sci USA **104**, 18976 (2007).

[3]    M. S. Cheung, D. Klimov, and D. Thirumalai, Proc Natl Acad Sci USA **102**, 4753 (2005).

[4]    D. Homouz, M. Perham, A. Samiotakis, M. S. Cheung, and P. Wittung-Stafshede, Proc Natl Acad Sci USA **105**, 11754 (2008).

[5]    T. Veitshans, D. Klimov, and D. Thirumalai, Fold. Des. **2**, 1 (1997).

[6]    M. R. Betancourt and D. Thirumalai, Protein Science **8**, 361 (1999).